\documentclass[twocolumn,showpacs,preprintnumbers]{revtex4}
\usepackage{amsmath}
\usepackage{graphicx}
\usepackage{dcolumn}
\usepackage{bm}

\begin{document}

\title{Quantum Repeaters based on Deterministic Storage of a Single Photon in distant Atomic Ensembles}
\author{D.Aghamalyan,$^{1}$ and Yu.Malakyan$^{1,2}$}
\email{yumal@ipr.sci.am} \affiliation{$^{1}$ Institute for
Physical Research, Armenian National Academy of Sciences,
Ashtarak-2, 0203, Armenia}
\affiliation{$^{2}$ Centre of Strong
Field Physics, Yerevan State University, 1 A.Manukian St.,Yerevan,
0025, Armenia}
\date{\today }

\begin{abstract}
Quantum repeaters hold the promise to prevent the photon losses in communication channels. Most recently, the serious efforts have been applied to achieve scalable distribution of entanglement over long distances. However, the probabilistic nature of entanglement generation and realistic
quantum memory storage times make the implementation of quantum repeaters an outstanding experimental challenge. We propose a quantum repeater protocol based on the deterministic storage of a single photon in atomic ensembles confined in distant high-finesse cavities and show that this system is capable of distributing the entanglement over long distances with a much higher rate as compared to previous protocols, thereby alleviating the limitations on the quantum memory lifetime by several orders of magnitude. Our scheme is robust with respect to phase
fluctuations in the quantum channel, while the fidelity imperfection is fixed and negligibly small at each step of entanglement swapping.
\end{abstract}

\maketitle




\section{\protect\normalsize INTRODUCTION}

Distribution of entanglement between distant matter nodes of
quantum networks is a challenging task because of the exponential
loss of photons in communication channels. One way to cover large
distances is using quantum repeater systems \cite{brieg}, which
split the quantum communication line into small segments and
combine local atomic memories for photons with entanglement
swapping to extend the entanglement generated between pairs of
neighboring memory elements over the entire communication link
length. Currently, a number of experiments toward realization of
scalable quantum repeater systems have been successfully
accomplished \cite{mats,cho,chou,laurat} on the basis of
the well-known Duan, Lukin, Cirac and Zoller (DLCZ)
protocol \cite{duan}, where it is proposed at first to generate and store
entanglement between two atomic ensembles and then to connect two
pairs of entangled ensembles. As an initial step, heralded
entanglement between two remotely located atomic ensembles is
established by detecting a single Stokes photon emitted
indistinguishably from either of the two ensembles via
spontaneous Raman process  resulting in the creation and storage
of collective spin excitation in atoms. This has been convincingly
demonstrated in the experiments mentioned above. However, the entanglement
connection between two pairs of entangled ensembles, which requires
controllable conversion of stored atomic excitations into
anti-Stokes photons followed by their detection in the same way as
the Stokes photons, was faced with serious difficulties
\cite{laurat}. The main reason is the probabilistic nature of the DLCZ
scheme based on the key requirement of low probability for
Stokes-photon emission that is needed to avoid contamination of
the entangled state by processes involving more than one atomic
excitation. As a result, the entanglement creation is achieved
only after many unsuccessful attempts that severely limit the
efficiency of entanglement swapping even under ideal conditions of
reconversion and detection of anti-Stokes photons. To overcome
this limitation new schemes for improvements of the DLCZ protocol
have recently been proposed \cite{san,choi,brask,simon,sang},
promising an exciting possibility for robust and efficient
entanglement generation. However, the predicted times for overall entanglement distribution
at large distances are still very long as compared to realistic
quantum memory storage times.
\begin{figure}[b] \rotatebox{0}{\includegraphics*
[scale = 0.55]{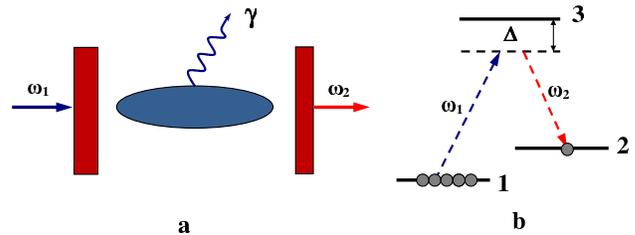}} \caption{(Color online)(a) Building block
with atomic ensemble confined in a microcavity. $\gamma$ shows
fluorescence photons emitted outside the cavity. (b) Atomic level
structure for emission of Stokes photon in far off-resonant Raman
configuration.}
\end{figure}

In this paper we propose a quantum repeater protocol based on a
deterministic storage of single photons in remote atomic ensembles, thus removing the inherent drawback
of the probabilistic DLCZ protocol. We show that our scheme is able to generate a single
cavity-mode Stokes photon with near-unit probability that provides very fast and robust entanglement swapping over large distances.
This advance is made possible by introducing two key modifications over existing protocols.
First, we employ a single-photon excitation of the ensembles to exclude the multiatom
events in the collective spin excitation and second, we use an ensemble of cold atoms strongly
coupled to a high-finesse cavity field that maximally enhances the Stokes-photon generation into the cavity mode.
The building block in our scheme is a Bose-Einstein condensate (BEC) of $N$ atoms with
$\Lambda$-type level structure, each of which is identically
and strongly coupled to the cavity mode [Fig.1(a)]. In contrast to the thermal atoms, where the individual, position-dependent
coupling for each atom to the cavity field leads to spatial inhomogeneities, in the BEC the atomic motion is almost frozen which allows one to maximize the collective coupling between the ensemble and the cavity field and to minimize position fluctuations keeping at the same time the number of atoms fixed. This has been recently realized experimentally in Refs. \cite{bren,colom}.

In Fig. 1, a single-photon at the frequency $\omega_1$ entering
the one-side cavity, for example, through the left mirror, which is assumed transparent for light at this frequency, excites the atoms in the
$|1\rangle \rightarrow |3\rangle$ transition and is converted via Raman
scattering into the cavity-mode $\omega_2$ photon [Fig.1(b)], which leaves the cavity through the right mirror with a transmissivity incomparably
larger than that of the left mirror. The result is the storage of incident photon in the medium as a single spin excitation, which is subsequently retrieved in the anti-Stokes photon. Both processes are deterministic thanks to the multiatom collective interference
effect and the cavity enhanced atom-light interaction. If all atoms are initially prepared in the
ground state $|0_a\rangle=|1_{1},..1_{i},....1_{N}\rangle$, then
upon emitting one cavity-mode Stokes photon the atomic ensemble
settles down into the symmetric state
\begin{equation}\label{1}
|1_a\rangle=S^{+}|0_a\rangle=\frac{1}{\sqrt{N}}\underset{i=1}{\overset{N}{\sum }}\mid 1_{1},..2_{i},....1_{N}\rangle
\end{equation}
with one spin excitation. The collective atomic spin operator is defined as
\begin{equation}\label{3}
S^{+}= \frac {1}{\sqrt{N}}\underset{j=1}{\overset{N}{\sum }}\sigma
_{21}^{(j)},\ \ \ \ \ \ \ \ \ \ \ S=(S^{+})^{\dagger}\ \
\end{equation}
obeying the commutation relation $\left[S,S^{+}\right]\simeq1$,
following from the fact that upon interacting with the
single photon, almost all the atoms are maintained in the ground
state $\left|1\right\rangle$. Here $\sigma _{\alpha \beta
}^{(j)}=\mid \alpha \rangle $ $_{j}$\ $\langle \beta \mid $\ is
the atomic spin-flip operator in the basis of two ground
states $\mid 1\rangle $\  and $\mid 2\rangle $ \ for the $j$th
atom. Through strongly suppressed Raman scattering
into other optical modes (see below), the output state of atoms and
photons can be written as
\begin{equation}\label{1}
|\Psi_{out}\rangle=\sqrt{1-p} \ |0_a\rangle \otimes|1_{\omega_1}\rangle
|0_{\omega_2}\rangle + \sqrt{p} \ |1_a\rangle
\otimes|0_{\omega_1}\rangle |1_{\omega_2}\rangle
\end{equation}
where $|0_{\omega_1}\rangle,|1_{\omega_1}\rangle$  and
$|0_{\omega_2}\rangle,|1_{\omega_2}\rangle$  denote Fock states
with zero and one photon of the incident and cavity fields,
respectively, and $p$ is the probability of cavity photon emission
by the atoms illuminated by the input single-photon pulse. The cavity mode is assumed to be quasi-cylindrical, while the atomic ensemble
is pencil shaped and is optically thick along the cavity axis.
As is shown below, this scheme is capable of producing cavity
photons with probability $p\sim1$, even if the one-photon detuning
$\Delta$ is kept much larger: $ \Delta \gg k, g_1, g_2$, which is
desirable to make the system robust against the spontaneous loss
from upper level and dephasing effects induced by other excited
states. Here $g_1$ and $g_2$ are the atom-field coupling constants
in the transitions $|1\rangle \rightarrow |3\rangle$ and
$|2\rangle \rightarrow |3\rangle$, respectively, and $k$ is the
cavity decay rate. This significant enhancement of atom-light
interaction is achieved due to the coherent coupling of different
atoms to the forward-scattered Stokes photon that creates a
collective atomic spin wave, which is in strong correlation with
the cavity mode. Meanwhile, the field modes other than the cavity
mode are weakly correlated with the collective atomic state and
contribute to noise resulting in the collectively enhanced
signal-to-noise ratio. This is also a basic property of the
original DLCZ protocol, where, however, instead of the input
single-photon pulse, a  short-pulse write laser is used which
gives rise to the problem of multiatom excitations.
Another distinctive feature of our approach is the
atom-cavity interaction in the regime of strong coupling:
$g_{1,2}\gg k$, which allows one to essentially amplify the
Stokes-photon generation.
\begin{figure}[b] \rotatebox{0}{\includegraphics*
[scale = 0.65]{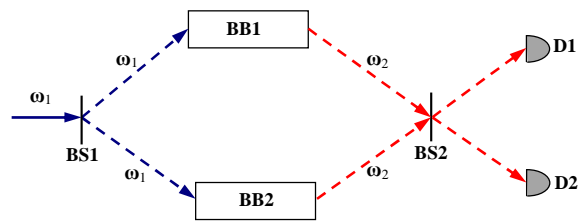}} \caption{(Color online) Set-up for
entanglement generation between two remote ensembles, which are
confined in the building blocks BB1 and BB2 shown in Fig. 1(a). A single
$\omega_1$-photon pulse passes through a 50:50 beam-splitter (BS1)
into two ensembles. The $\omega_2$ Stokes photon emitted by
either of the ensembles is mixed on a balanced beam-splitter (BS2)
and detected at single photon detectors D1 and D2, thus projecting
the ensembles into an entangled state with one spin excitation   stored in either  of the ensembles.
The $\omega_1$ photon source and detectors D1 and D2 are located in a
central station between the ensembles. }
\end{figure}

The procedure for entanglement creation between the
remote locations BB1 and BB2 in Fig.2 requires that either of the two atomic ensembles is excited by the $\omega_1$-photon such that a single Stokes photon is emitted from two ensembles. The detection of the Stokes photon, for example in D1, projects
the joint state of atomic ensembles in an entangled state having the form
\begin{equation}
|\Psi_{12}\rangle=\frac {1}{\sqrt{2}}( |1_a\rangle_1 |0_a\rangle_2 + \exp(i\Phi_{12})|0_a\rangle_1 |1_a\rangle_2),
\end{equation}
where $|0_a\rangle_i$ and $|1_a\rangle_i$ are the atomic collective states for the $i$th ensemble ($i$=1,2) with 0 and 1 spin excitation, respectively, and $\Phi_{12}=\phi_{\omega_1}^{(1)}+\phi_{St}^{(1)}-(\phi_{\omega_1}^{(2)}+\phi_{St}^{(2)})$ stands for the difference of phases $\phi_{\omega_1,St}^{(1)}$ and $\phi_{\omega_1,St}^{(2)}$  acquired by the $\omega_1$- and Stokes photons on their way in the up [labeled as (1)] and down (2) sides of the channel. The state (4) is generated with efficiency increasing linearly with the Stokes-photon emission probability $p$, but the entanglement creation process remains probabilistic due to photon losses and various imperfections in the quantum channel, as is shown in Sec.III. Note, however, that although the suppression of photon losses in communication channels is still a serious experimental challenge, we hope that this problem will be solved by taking advantage of hollow-core photonic-crystal fibers \cite{russel}. Then, taking also into account the recent progress achieved to highly enhance the efficiencies of single-photon detection \cite{james} and of retrieving anti-Stokes photons in cavities \cite{sim}, our scheme will provide a deterministic entanglement swapping in every step with a success probability close to unity that is a fundamental difference from previous quantum repeater protocols, which even in the hypothetical case of zero losses cannot be deterministic, as their functionality relies crucially on the key requirement of low probability of Stokes-photon emission.

This paper is organized as follows. In the next section we show the
ability of the presented scheme to produce cavity photons with probability $p\sim1$.
Here we find the analytic solutions for the flux and
number of output Stokes photons and prove the generation of the state (3).
Then, in Sec. III we calculate the entanglement distribution time taking into account different losses and imperfections in the entanglement swapping and demonstrate that our protocol is the fastest one and is robust with respect to the pathway phase fluctuations.
Our conclusions are summarized in Sec. V.

\section{\protect\normalsize DETERMINISTIC GENERATION OF CAVITY PHOTONS}

In what follows, we use the approach employed in our earlier work
\cite{sis}, where the quantum theory for generation of correlated
Stokes and anti-Stokes photons in the DLCZ protocol is developed. We
start from the effective Hamiltonian of the system in the rotating frame
\begin{equation}\label{2}
H=\hbar G \sqrt{N}\left[S^{+}a_{2}^{+}a_{1}+H.c\right],
\end{equation}
which is obtained by adiabatically eliminating the upper state
$\mid 3\rangle $ owing to the large one-photon detuning $\Delta
=\omega _{31}-\omega _1=\omega _{32}-\omega _2$. In Eq.(5),
$a_i(a_i^{+})$ is the annihilation (creation) operator of $i$th
field inside the cavity, $G=g_1g_2/\Delta$, and
$g_i=\sqrt{{2\pi\omega_i} /{\hbar V}}\mu_{3\alpha}$, with $V$ being the
quantization volume taken equal to the interaction volume and
$\mu_{\alpha \beta}$ being the dipole matrix element of the
 $\left|\alpha\right\rangle\rightarrow\left|\beta\right\rangle$ transition.

With the Hamiltonian (5), the Heisenberg-Langevin equations for
$a_i(t)$ are given by \cite{gard}
\begin{equation}\label{4}
\dot{a_1}(t)=-iG \sqrt{N}S a_{2}(t)-\frac{1}{2}(\chi+\Gamma
N)a_{1}(t)-\sqrt{\chi}a_{1,\text{in}}(t)
\end{equation}
\begin{equation}\label{5}
\dot{a_2}(t)=-iG \sqrt{N}S^{+} a_{1}(t)-\frac{1}{2}k
a_{2}(t)-\sqrt{k}a_{2,\text{in}}(t),
\end{equation}
where $a_{i,in}(t)$ are the operators of input fields with the
properties $[a_{i,\text{in}}(t),a_{j,\text{in}}^{+}(t^{\prime })]=\delta_{ij} \delta
(t-t^{\prime })$, $\chi$ is the photon number damping rate for the
$\omega_1$ field, which is calculated in the free-space limit as
the inverse of propagation time of the $\omega_1$ pulse through
the atomic sample: $\chi=c/L$, with $L$ being the sample length. In
Eq. (6) we have included also the losses $\Gamma N$ of
the $\omega_1$ field, which originate from the emission of fluorescent photons outside
the cavity [Fig. 1(a)], where $\Gamma=(g_1/\Delta)^2 \gamma_3$
\cite{sis} ($\gamma_3$ is the spontaneous decay rate of the upper
level 3). However, these losses are usually very small as compared to
$\chi$, $\Gamma N\ll \chi$, and are neglected below.

For an input $\omega_1$ pulse with duration $T\gg (\chi,k)^{-1}$,
the solution of Eqs. (6) and (7) at large times takes the form
\begin{equation}\label{6}
a_1(t)=-{\frac{2}{\sqrt \chi}}\frac{1}{1+\eta
SS^{+}}(a_{1,\text{in}}(t)-i\sqrt\eta Sa_{2,\text{in}}(t)),
\end{equation}
\begin{equation}\label{7}
a_2(t)=-{\frac{2}{\sqrt k}}\frac{1}{1+\eta
S^{+}S}(a_{2,\text{in}}(t)-i\sqrt\eta S^{+}a_{1,\text{in}}(t)),
\end{equation}
where $\eta=4NG^2/(\chi k)$ represents the product of the coherent interaction rate $4NG^2/k$ with interaction time $\chi^{-1}$.

We introduce the output photon operators $\hat{n}_{i,\text{out}}(t)=\int\limits_{-\infty}^{t}
a^{\dag}_{i,\text{out}}(\tau) a_{i,\text{out}}(\tau)d\tau$
and find their mean values $n_{i,\text{out}}(t)=\langle\hat{n}_{i,\text{out}}(t)\rangle$ from the flux equations
\cite{blow}
\begin{equation}\label{8}
\frac{dn_{i,\text{out}}(t)}{dt} = \langle
a^{\dag}_{i,\text{out}}(t) a_{i,\text{out}}(t)\rangle, \ \ i=1,2,
\end{equation}
where the output fields $a_{i,\text{out}}(t)$ are connected with the input
$a_{i,\text{in}}(t)$ and intracavity $a_i(t)$ fields by the
input-output formalism \cite{gard}
\begin{equation}\label{9}
a_{i,\text{out}}(t)- a_{i,\text{in}}(t)=\sqrt{r_i}a_i(t)\ \ i=1,2,
\end{equation}
with $r_{1,2}=(\chi,k)$. The mean value $\langle
\hat{O}\rangle=\langle\Psi_\text{in}\mid \hat{O}\mid\Psi_\text{in}\rangle$
of any Heisenberg operator $\hat{O}$ is calculated with the initial
state $|\Psi_{\text{in}}\rangle=|0_a\rangle \otimes|1_{\omega_1}\rangle
|0_{\omega_2}\rangle$. Then, using Eqs. (8), (9) and (11) and recalling that
$SS^{+}|\Psi_{\text{in}}\rangle\simeq |\Psi_{\text{in}}\rangle$, for the
vacuum input at Stokes frequency $\langle a^{\dag}_{\text{2,in}}(t) a_{\text{2,in}}(t)\rangle = 0$ we readily find
\begin{equation}\label{10}
\frac{dn_{1,\text{out}}(t)}{dt} =
\frac{(1-\eta)^2}{(1+\eta)^2}\mid f(t) \mid^2
\end{equation}
\begin{equation}\label{11}
\frac{dn_{2,\text{out}}(t)}{dt} = \frac{4 \eta}{(1+\eta)^2}\mid
f(t) \mid^2
\end{equation}
where the initial flux $\langle
a^{\dag}_{1,\text{in}}(t) a_{1,\text{in}}(t)\rangle$ of $\omega_1$ field is expressed
in terms of the pulse temporal envelope $f(t)$, given by
\begin{equation}\label{12}
\langle 0\mid a_{1,\text{in}}(t)\mid 1_{\omega_1}\rangle =f(t)
\end{equation}
and normalized as $\int\limits_{-\infty}^{\infty} \mid f(t)\mid^2
dt=1$, indicating that the number of impinged photons is 1. Similarly, the wave functions of the output modes are determined as
$\langle 0\mid a_{i,\text{out}}(t)\mid \Psi_{\text{in}}\rangle =\Phi_i(t)$, giving for the output photon numbers
\begin{equation}\label{12}
n_{i,\text{out}}\equiv n_{i,\text{out}}(\infty)=\int\limits_{-\infty}^{\infty}\mid \Phi_i(t')\mid^2dt'.
\end{equation}

From Eqs. (12)-(14) it follows that the waveform of the emitted
Stokes-photon reproduces the shape of the
input $\omega_1$ pulse up to a constant factor. Besides, the total number of photons is
conserved: $n_{1,\text{out}}(\infty)+n_{2,\text{out}}(\infty)=n_{1,\text{in}}=1$,
showing the ability of the system to produce Stokes photons with
probability $p=n_{2,\text{out}}/n_{1,\text{in}}=1$, if
$\eta=1$ or
\begin{equation}\label{13}
\frac{4NG^2}{k}=\chi.
\end{equation}
Thus, we arrive at a quite reasonable requirement that for deterministic Stokes-photon generation the collectively enhanced Raman process should be
as fast as the passage of the $\omega_1$ pulse through the atomic
sample. In this case the Stokes photon wave form (13) is identical to the input pulse shape $\mid f(t) \mid^2$.
Note, that for larger values of $\eta$ the conversion is not
complete, $n_{1,\text{out}}(\infty)>0$, because of, although weak, backward transformation of the
Stokes photon into the $\omega_1$ photon.

It is useful to consider numerical estimations at this point.
As a sample the $^{87}Rb$ vapor is chosen with the ground states
$5S_{1/2}(F=1)$ and $ 5S_{1/2}(F=2)$ and the excited state
$5P_{3/2}(F^{'}=2)$ being the atomic states $1, 2$ and  $3$ in
Fig. 1(b), respectively. For the light wavelength $\lambda \simeq 0.8\mu
$m, $\gamma_3 =2\pi \times 6$ MHz, $g_1\sim g_2=10\gamma_3,\Delta
\sim 50\gamma_3, k=3\gamma_3$, and atomic trap length $L\sim 100
\mu $m we find that Eq.(14) is fulfilled with $N\sim 10^6$. At the same time the number of fluorescent photons is
negligibly small: $\Gamma N/\chi\ll 1$. All
these parameters appear to be within experimental reach, including the deterministic sources of
initial narrow-band single-photon pulses with a duration
of several microseconds \cite{kuhn,chen,mckeever,hijl,gog} and
BEC with $\sim 10^6$ atoms in the QED cavity \cite{bren,colom}.

In the Schrodinger picture, we obtain the output state $\mid\psi_{out}\rangle$ of the system as the eigenstate of
total photon number operator
\begin{equation}
(\hat n_{1,\text{out}}(\infty)+\hat n_{2,\text{out}}(\infty))\mid \psi_{\text{out}}\rangle=\mid
\psi_{\text{out}}\rangle
\end{equation}
where the output modes are described by known wave functions $\Phi_i(t)$. To construct this state, we introduce the operators of creation of single-photon wave packets at frequencies $\omega_{i}$ associated with mode functions $\Phi_{i}(t)$ as \cite{gogyan}
\begin{equation}
\hat c_i^{\dag}=\frac{1}{\sqrt{n_{i,\text{out}}}}\int \limits_{-\infty}^{\infty}dt \Phi_i(t)a^{\dag}_{i,\text{out}}(t)
\end{equation}
These operators create single-photon states in the usual way
by acting on the field vacuum $\mid 0_f\rangle=\prod_{i}\mid 0_{\omega_i}\rangle$
\begin{equation}
\hat c_{i}^{\dagger}\mid 0_f\rangle=\mid 1_{\omega_i}\rangle \mid 0_{\omega_j}\rangle, \ \ \ i,j=1,2;\ \ i\neq j
\end{equation}
where Eq. (15) has been used. In the right-hand side of Eq. (19) the field vacuum is reduced to $\mid 0_f\rangle=\mid 0_{\omega_1}\rangle\mid 0_{\omega_2}\rangle$, since other modes are not occupied by the photons and, hence, are not taken into account during the measurements. The operators $c_i$ satisfy the standard boson commutation relations,
\begin{equation}
[\hat c_{i},\hat c_{j}^{+}]=\delta_{ij},
\end{equation}
that follow from the commutation relations of the output fields $\langle [a_{i,\text{out}}(t), a^{\dag}_{j,\text{out}}(t')]\rangle=\delta_{ij}\delta (t-t')$, which are simply found using Eqs. (8), (9) and (11). From this, we easily find the output state
\begin{equation}
|\Psi_{\text{out}}\rangle=\sqrt{n_{1,\text{out}}}\ |0_a\rangle \otimes \ c^{\dag}_1|0_f\rangle + \sqrt{n_{2,\text{out}}}\ |1_a\rangle \otimes c^{\dag}_2|0_f\rangle,
\end{equation}
which is exactly the state (3). In the general case of $n_{1,2,\text{out}}\neq 0$, the system produces a
photonic qubit, i.e., a single-photon state entangled in two
distinct frequency modes with wave functions
$\Phi_{1}(t)$ and $\Phi_{2}(t)$. The pure output state $|\psi_{\text{out}}\rangle=|1_a\rangle \otimes |0_{\omega_1}\rangle |1_{\omega_2}\rangle$ consisting of a single Stokes photon and one spin excitation in the atomic ensemble is obtained in the limit of complete conversion of input $\omega_{1}$ mode into the Stokes photon under the condition (16). From Eq. (21) we immediately find the number of spin-wave excitations $N_{\text{sp}}=\langle S^{+}S \rangle= n_{2,\text{out}}$. This result is followed also from the comparison of Heisenberg equations for $\hat N_{\text{sp}}$ and $\hat n_{2,\text{out}}$, if the relaxations are negligibly small, as is the case here.

\section{\protect\normalsize ENTANGLEMENT DISTRIBUTION RATE}

The time $T_{\text{tot}}$ required for entanglement distribution over a distance $L$ can be calculated by the same methods that are used for the original DLCZ protocol \cite{duan,sang}. The main difference between the two schemes occurs only in the entanglement preparation stage, so that, by separating out the success probability $P_0$ of entanglement generation between two atomic ensembles in the elementary link, the general formula (13) for the distribution time obtained in Ref.\cite{sang} can be rewritten as
\begin{equation}
T_{\text{tot}}= T_{\text{tot,DLCZ}}\frac {P_{0,\text{DLCZ}}}{P_0}.
\end{equation}
Here we have taken into account that in the entanglement swapping the anti-Stokes photon being far from the resonance with the cavity is generated as
in the free space and, hence, the cavity has no effect on the excitation
transfer from the collective atomic mode to the optical mode.
The dominant noise that limits the success probability $P_0$ is the photon losses,
which include the transmission channel losses and the inefficiency of single-photon detectors.
Correspondingly, $P_0=p\eta_d \eta_t$ is defined as the product of quantum-mechanical probability $p$
of Stokes-photon emission by the photon detection efficiency $\eta_d$ and the transmission
efficiency $\eta_t=\text{exp}(-\frac{L_0}{L_{\text{att}}})$,
where $L_0$ is the distance between the atomic ensembles (or the length of the elementary link)
in Fig. 2 and $L_{\text{att}}$ is the communication channel attenuation length. Compared to the DLCZ scheme, the transmission
efficiency is quadratically smaller: $\eta_t=\eta_{t,\text{DLCZ}}^2$, since in our case we have
to include the transmission losses both for $\omega_1$ photon propagating from the source to
the atomic ensembles and for the Stokes photon propagating back to the detectors placed in the central station. This leads to a notable increase of $T_{\text{tot}}$, which, however, is compensated by another effect. Indeed, the deterministic generation of the Stokes photon relaxes the limitations for the quantum memory lifetime, thus allowing one to enhance the memory efficiency $\eta_m$. Since $T_{\text{tot, DLCZ}}$ in Eq.(22) is inverse proportional to the success probabilities at each level of entanglement connection or to $\eta_m^{n+2}$ \cite{sang}, where $L_0=\frac{L}{2^n}$, then by properly choosing the parameters one can ensure that the two effects of lowering $\eta_t$ and increasing of $\eta_m$ cancel each other in $T_{\text{tot}}$ for arbitrary communication length $L$. As such, taking into account that usually $p_{\text{DLCZ}}< 0.01$, we find that in the present scheme with $p\sim1$ the entanglement distribution rate $T_{\text{tot}}^{-1}$ increases at least by 2 orders of magnitude as compared to that of the original DLCZ protocol. For example, for the parameters used in Fig. 18 of Ref. \cite{sang} $L_{\text{att}}\sim $22km (this corresponds to a fiber attenuation of 0.2 dB/km for telecom wavelength photons), $\eta_d \sim 1$ \cite{james}, and $n=4$, and for $\eta_m\sim 1$, the total time needed in our scheme for distributing a single entangled pair over the distance $L = 1000$ km is only 10 s, thus making our protocol the fastest one and comparable with the multimode-memory-based protocol of Ref. \cite{gisin}. It is worth noting that, while achieving much faster performance, our proposal has an advantage in robustness. In the probabilistic protocols \cite{sang}, the errors which reduce the fidelity of the distributed state are mainly caused by the event when more than one atom is excited into the collective spin wave, whereas only one Stokes photon is detected. This process is forbidden in our scheme, and, hence, the fidelity imperfection is fixed and negligibly small during all the time of entanglement distribution. Furthermore, our scheme is much more robust with respect to phase fluctuations in the fibers. To show this let us remind that the phase $\Phi_{12}$ in Eq.(4) is sensitive to path-length fluctuations leading to phase instability of entanglement connection between the two pairs of entangled ensembles \cite{duan,sang}. Suppose that two pairs of atomic ensembles (1,2) and (3,4) are prepared in independent entangled states like the state (4) with the phases $\Phi_{12}$ and $\Phi_{34}$, respectively, and that the connection between the pairs performed by the standard procedure \cite{duan,sang} projects their state into a maximally entangled  state between the four atomic ensembles,
\begin{equation}
|\Psi_{12,34}\rangle=\frac {1}{\sqrt{2}}( |1_a\rangle_1 |1_a\rangle_4 + \exp(i\delta\Phi)|1_a\rangle_3 |1_a\rangle_2),
\end{equation}
where $\delta\Phi=\Phi_{12}-\Phi_{34}$ is the relative phase between the two entangled pairs. Since the entanglement generation process is probabilistic and, hence it is established in the pairs at different time moments although the entanglement preparation begins in the two pairs simultaneously, the phases $\Phi_{12}$ and $\Phi_{34}$ are different due to path length fluctuations during this time interval. The mean value of the latter can be estimated as the duration of entanglement generation between two atomic ensembles in the elementary link: $\tau\sim L_0/(cP_0)$. For $L_0\sim 60$km, $p=1$ and for the rest parameters given above we have $\tau\sim 0.004$s. This means that the Stokes-photon coming time must be controlled over this averaged time with accuracy of the order of $\omega^{-1}\sim 10^{-15}$s ($\omega\sim \omega_1\sim\omega_2$),  corresponding to $\delta \Phi<1$, which is achievable for current technologies \cite{hol}. Note that in the DLCZ protocol $\tau\sim 1$s for the same parameters that makes the phase stabilization hardly feasible. A much more favorable situation occurs if the both the entanglement generation and the swapping are done via two-photon detection similar to the protocol proposed in Ref. \cite{pan}, in contrast to the DLCZ protocol based on a single-photon detection. As has been shown in Ref. \cite{pan}, in this case the propagation phases that two photons acquire only lead to a multiplicative factor to the pair entangled state. Moreover, if in Ref. \cite{pan} the total state-function of two entangled pairs, apart from the entangled part, contains also the contribution from multiatom excitations that deteriorates the final-state fidelity, in our case the photonic part of the state (3) for $p=1$ has no vacuum and two- or more photon components; hence the two pairs are in a pure maximally entangled state, thus making the long-distance phase stabilization unnecessary.

\section{\protect\normalsize CONCLUSIONS}

Summarizing, we have described a quantum repeater protocol that uses an input single-photon pulse instead of a write laser in the original DLCZ scheme, thus avoiding the problem of multiple atomic spin excitations. Also a high-finesse cavity is employed to maximally enhance the Raman conversion of input photon into forward scattered Stokes light mode. The main advantage of this setup is that it does not constrain the probability of Stokes-photon generation, which can be made equal to unity by adjusting the system parameters. As a result, the errors, which reduce the fidelity in the conventional DLCZ protocol, are strongly suppressed and the long-distance interferometric stability is no longer required. Our scheme enables a robust quantum repeater without long-living quantum memories, thus providing a fast communication rate. The proposed protocol can be also implemented with atoms confined inside a single-mode hollow-core photonic-crystal fiber.

\bigskip
\subsection*{Acknowledgments}

This research has been conducted in the scope of the International
Associated Laboratory IRMAS. We also acknowledge support from the
Science Basic Foundation of the Government of the Republic
of Armenia.



\bigskip

\begin{references}

\bibitem{brieg} H. -J. Briegel, W.Dur, J. I. Cirac, and P. Zoller, Phys.Rev.Lett. {\bf 81}, 5932 (1998).

\bibitem{mats} D. N. Matsukevich and A. Kuzmich, Science {\bf 306}, 663 (2004).

\bibitem{cho} C. W. Chou, H. de Riedmatten, D. Felinto, S. V. Polyakov, S. J. van Enk, and H. J. Kimble, Nature (London) {\bf 438}, 828 (2005).

\bibitem{chou} C.W.Chou, J. Laurat, H. Deng, K. Choi, H. de Riedmatten, D. Felinto, and H.J. Kimble, Science {\bf 316}, 1316 (2007).

\bibitem{laurat} J.Laurat, C.-W Chou, H. Deng, K. S. Choi, D. Felinto, H. de Riedmatten and H J Kimble, New J. Phys. {\bf 9}, 207 (2007).

\bibitem{duan} L. M. Duan, M. D. Lukin, J. I.Cirac, and P. Zoller, Nature (London) {\bf 414}, 413 (2001).

\bibitem{san} N. Sangouard, C. Simon, J. Minar, H. Zbinden, H. de Riedmatten, and N. Gisin, Phys. Rev. A {\bf 76}, 050301(R)(2007).

\bibitem{choi} K. S. Choi, H. Deng, J. Laurat, and H. J. Kimble, Nature (London) {\bf 452}, 67 (2008).

\bibitem{brask} J. B. Brask, L. Jiang, A. V. Gorshkov, V. Vuletic, A. S. Sorensen, and M. D. Lukin, Phys. Rev. A {\bf 81}, 020303(R)(2010).

\bibitem{simon} C. Simon, H. de Riedmatten and M. Afzelius, Phys. Rev. A {\bf 82}, 010304(R) (2010).

\bibitem{sang} N. Sangouard, C. Simon, H. de Riedmatten, and N. Gisin, Rev. Mod. Phys. {\bf 83}, 33 (2011).

\bibitem{bren} F.Brennecke, T. Donner, S. Ritter, T. Bourdel, M. Kohl, and T.Esslinger, Nature (London) {\bf 450}, 268 (2007).

\bibitem{colom} Y. Colombe, T. Steinmetz, G. Dubois, F. Linke, D. Hunger, and J. Reichel, Nature (London) {\bf 450}, 272 (2007).

\bibitem{russel} P. St. J. Russell, J. Lightwave Technol. {\bf 24}, 4729 (2006).

\bibitem{james} D.F.V. James and P.G. Kwiat, Phys. Rev. Lett. {\bf 89}, 183601 (2002); A. Imamoglu, ibid. {\bf 89}, 163602 (2002).

\bibitem{sim} J.Simon, H. Tanji, J. K. Thompson, and V. Vuletic, Phys. Rev. Lett. {\bf 98}, 183601 (2007).

\bibitem{sis} N. Sisakyan and Yu.Malakyan, Phys.Rev. A {\bf 72}, 043806 (2005).

\bibitem{gard} C. W. Gardiner and P. Zoller, {\it Quantum Noise}(Springer-Verlag, Berlin, 1999).

\bibitem{blow} K.J.Blow, R.Loudon, S.J.D.Phoenix, and T.J.Shepherd, Phys. Rev. A {\bf 42}, 4102 (1990).

\bibitem{kuhn} A. Kuhn, M. Hennrich, and G. Rempe, Phys.Rev.Lett. {\bf 89}, 067901 (2002).

\bibitem{chen} S. Chen, Yu-Ao Chen, T.Strassel, Z.-S. Yuan, Bo Zhao, J.Schmiedmayer, and J.-W. Pan, Phys.Rev.Lett. {\bf 97}, 173004 (2006).

\bibitem{mckeever} J. McKeever, A. Boca, A. D. Boozer, R. Miller, J. R. Buck, A. Kuzmich, H. J. Kimble, Science {\bf 303}, 1992 (2004).

\bibitem{hijl} M. Hijlkema, B. Weber, H. P. Specht, S. C. Webster, A. Kuhn, and G. Rempe, Nature Phys. {\bf 3}, 253 (2007).

\bibitem{gog} A. Gogyan, S. Guerin, H.-R. Jauslin, and Yu. Malakyan, Phys. Rev. A, {\bf 82}, 023821 (2010).

\bibitem{gogyan} A.Gogyan and Yu.Malakyan, Phys.Rev. A {\bf 77},  033822 (2008).

\bibitem{gisin} C.Simon, H. de Riedmatten, M. Afzelius, N. Sangouard, H. Zbinden, and N. Gisin, Phys. Rev. Lett. {\bf 98}, 190503 (2007).

\bibitem{hol} K. W. Holman, D. D. Hudson, and J. Ye, Opt. Lett. 30, 1225 (2005).

\bibitem{pan} Z.-B. Chen, Bo Zhao, Yu-Ao Chen, J. Schmiedmayer, and J.-W. Pan, Phys.Rev. A {\bf 76},  022329 (2007).



\end{references}
\end{document}